\documentstyle[12pt]{article}
\catcode`@=11
\newcount\@tempcntc
\def\@citex[#1]#2{\if@filesw\immediate\write\@auxout{\string\citation{#2}}\fi
  \@tempcnta\z@\@tempcntb\m@ne\def\@citea{}\@cite{\@for\@citeb:=#2\do
    {\@ifundefined
       {b@\@citeb}{\@citeo\@tempcntb\m@ne\@citea\def\@citea{,}{\bf ?}\@warning
       {Citation `\@citeb' on page \thepage \space undefined}}%
    {\setbox\z@\hbox{\global\@tempcntc0\csname b@\@citeb\endcsname\relax}%
     \ifnum\@tempcntc=\z@ \@citeo\@tempcntb\m@ne
       \@citea\def\@citea{,}\hbox{\csname b@\@citeb\endcsname}%
     \else
      \advance\@tempcntb\@ne
      \ifnum\@tempcntb=\@tempcntc
      \else\advance\@tempcntb\m@ne\@citeo
      \@tempcnta\@tempcntc\@tempcntb\@tempcntc\fi\fi}}\@citeo}{#1}}
\def\@citeo{\ifnum\@tempcnta>\@tempcntb\else\@citea\def\@citea{,}%
  \ifnum\@tempcnta=\@tempcntb\the\@tempcnta\else
   {\advance\@tempcnta\@ne\ifnum\@tempcnta=\@tempcntb \else \def\@citea{--}\fi
    \advance\@tempcnta\m@ne\the\@tempcnta\@citea\the\@tempcntb}\fi\fi}
\catcode`@=12
\def\barr{\begin{array}}
\def\earr{\end{array}}
\def\bea{\begin{eqnarray}}
\def\eea{\end{eqnarray}}
\def\bmath{\begin{displaymath}}
\def\emath{\end{displaymath}}
\def\bq{\begin{quote}}
\def\eq{\end{quote}}
\def\slash#1{\setbox0=\hbox{$#1$}#1\hskip-\wd0\hbox to\wd0{\hss\sl/\/\hss}}

\def\as{\alpha_s}

\def\CF{C_{\scriptscriptstyle F}}
\def\Li{\mbox{$\mbox{\rm Li}_2$}}
\def\Frac#1#2{\mbox{$\textstyle{#1\over#2}$}}

\def\be{\begin{equation}}
\def\ee{\end{equation}}
\def\bea{\begin{eqnarray}}
\def\eea{\end{eqnarray}}
\def\nn{\nonumber\\}

\def\as{\alpha_s}
\def\sw{s_{\scriptscriptstyle W}}

\def\real{\mathop{\mbox{\rm Re}}\nolimits}

\def\cz{\chi_{\scriptscriptstyle Z}}
\def\MZ{M_{\scriptscriptstyle Z}}
\def\GZ{\Gamma_{\scriptscriptstyle Z}}
\def\PL{P_{\scriptscriptstyle L}}
\def\I#1{\mbox{$I_{#1}$}}
\def\S#1{\mbox{$S_{#1}$}}
\begin{document}
\thispagestyle{empty}
\begin{flushright}
FTUV/94-4 \\[-0.2cm]
(revised version)
\end{flushright}
\begin{center}
{\bf{\Large $\boldmath O(\as)$ Longitudinal Spin Polarization}}\\[0.3cm]
{\bf{\Large in Heavy-Quark Production}} \\[2.5cm]
{\large Michael M.\ Tung}\footnote[1]{Electronic address:
tung@evalvx.if\/ic.uv.es} \\[0.4cm]
Departament de F\'\i sica Te\`orica, Universitat~de Val\`encia, and IFIC,\\
Universitat~de Val\`encia--CSIC, E-46100 Burjassot (Val\`encia), Spain.
\end{center}
\vspace{2.25cm}
\centerline {\bf ABSTRACT}
\noindent
We present the massive one-loop {\it QCD} corrections to the production cross
sections of polarized quarks in the annihilation process
$e^+e^-\to q\bar{q}(g)$ for bottom, top, and charm quarks.
{}From the full analytical expressions for the production cross sections,
Schwinger-type interpolation formulae for all parity-parity combinations
$(VV, V\!A, AA)$ are derived. The parity-odd interpolation formula contains the
correct limit for vanishing quark masses taking into account a residual
coupling
of left- and right-chiral states in the massless theory. Numerical results for
the total cross section and the longitudinal spin polarization demonstrate the
accuracy of the interpolation formulae. \\[1cm]
\noindent PACS number(s): 11.15.Bt, 12.38.Bx, 13.88.+e \\

\newpage
\noindent
Recent precision measurements on $Z$ decays by the {\it LEP} and {\it SLAC}
collaborations~\cite{LEP,SLAC} permit a thorough investigation of the
properties
of electroweak neutral currents and thus provide a stringent test of the
Standard
Model. Moreover, the study of polarized and unpolarized observables of the
annihilation process $e^+ e^- \to\gamma,Z\to q\bar{q}$ to the level of strong
radiative corrections offers an ideal experimental approach to confirm the
validity of quantum chromodynamics ({\it QCD\/}).

In the present letter, we concentrate on the effects that finite quark
masses have on the longitudinal polarization asymmetry $\PL$ of heavy quarks
produced through $e^+ e^-$ collisions~\cite{KPT}. We shall present a detailed
numerical
analysis of the $O(\as)$ corrections to the Born result for bottom, top, and
charm quarks and then compare these results with approximations stemming from
Schwinger-type interpolation formulae for the vector (V) and axial-vector (A)
combinations of the production cross sections. In particular, the interpolation
formulae for the parity-odd contribution $\sigma^{V\!A}_S$ have not been given
in
the literature before. They provide a valuable tool for expressing the $O(\as)$
longitudinal polarization at the one-loop order of strong interactions in a
condensed form.

For heavy-quark production the lowest order $\gamma$--$Z$ exchange cross
section
formulae can be written in terms of the different parity-parity combinations.
The parity-even property of the total unpolarized rate can directly be seen
from the decomposition~\cite{jers}
\be\label{1}
\sigma(e^+ e^-\to\gamma,Z\to q\bar{q}) =
\Frac{1}{2}v(3-v^2)\,\sigma^{VV}+v^3\,\sigma^{AA},
\ee
whereas the total parity-odd contribution relevant to the longitudinal
spin polarization of the final quark $q$ reads
\be\label{2}
\sigma(e^+ e^-\to\gamma,Z\to q(\lambda_\pm)\bar{q}) =
\pm v^2\,\sigma^{V\!A}_S.
\ee
Here, the two distinct helicity states of the quark are given by
$\lambda_\pm=\pm 1/2$. The velocity of the quark in the two-particle
final-state is $v=\sqrt{1-\xi}$ with $\xi=4m_q^2/s$, where as usual $\sqrt{s}$
is the center-of-momentum energy in the $e^+ e^-$ system and $m_q$ the mass
of the quark $q$. In Eqs.~(\ref{1}) and (\ref{2}) the vector, axial, and
interference contributions are explicitly given by
\bea\label{sigvv}
\sigma^{VV} &=& {4\pi\alpha^2\over s}\left[Q_q^2-2 Q_q v_e v_q \real{\cz}+
                (v^2_e+a^2_e) v_q^2 |\cz|^2\right], \\
\sigma^{AA} &=& {4\pi\alpha^2\over s}\,(v^2_e+a^2_e) a_q^2 |\cz|^2,
\eea
and
\be\label{sigva}
\hspace{-8mm}\sigma^{V\!A}_S\ =\ {4\pi\alpha^2\over s}\left[
              Q_q v_e a_q \real{\cz}-(v^2_e+a^2_e) v_q a_q |\cz|^2\right],
\ee
where $v_f=2T^f_z-4Q_f\sw^2$ and $a_f=2T^f_z$ are the electroweak vector- and
axial-vector coupling constants for fermions ($f$), respectively, and $Q_q$
denotes the fractional charge of the quark $q$. The quantity
$\cz(s)=g_{\scriptscriptstyle W} \MZ^2 s(s-\MZ^2+i\MZ\GZ)^{-1}$
characterizes the Breit-Wigner form of the $Z$ propagator, where $\MZ$ and
$\GZ$ are the mass and the total decay width of the $Z$ boson, and
$g_{\scriptscriptstyle W}=4.49\cdot 10^{-5}\,\mbox{\rm GeV}^{-2}$. Note that
we do not consider polarization of the initial beam, so that the axial-vector
part in $\sigma^{V\!A}_S$ stems from the spin projection operator in
the final state, and therefore $S$ denotes the linear dependence on the spin
of the final quark.

For extracting the physics of this process at the $Z$ it is necessary to
include
radiative corrections. Obviously, {\it QCD} corrections occur only in the final
state and will be proportional to $\as/\pi$, {\it i.e.\/}, approximately $4\%$,
whereas the electromagnetic final state corrections are only of the order of
$3\alpha Q_q^2/4\pi<0.1\%$ and can thus safely be neglected. Note that initial
state bremsstrahlung is significant and should be taken into
account~\cite{alex}.

To obtain the first-order {\it QCD} corrections for the individual
parity-parity contributions to the total annihilation rate, Eqs.\
(\ref{sigvv})--(\ref{sigva}), one has to include virtual-gluon exchange and
real-gluon emission. The virtual processes have the same final state
as the Born term and thus simply amount to a replacement of the $O(\as)$
massive vertex functions according to the substitutions
\be
\Gamma^V_\mu \,=\, (1+A)\gamma_\mu-B{(p_1-p_2)_\mu\over 2m_q},\quad
\Gamma^A_\mu \,=\, (1+C)\gamma_\mu\gamma_5+D{(p_1+p_2)_\mu\over 2m_q}\gamma_5,
\ee
where $A$, $B$, $C$, and $D$ are the so-called {\it chromomagnetic\/} form
factors~\cite{KPT}, and $p_1$ and $p_2$ are the momenta of quark and antiquark,
respectively (the superscripts $V$ and $A$ refer to the parity property
of the vertex function). However, in the one-loop corrections to
$\sigma^{V\!A}_S$ all terms proportional to $D$ will eventually vanish after
contracting the corresponding parity-odd hadron tensor with the usual lepton
tensor.

The kinematics of a three-body final state with a polarized quark is
considerably more complicated than the simple quark pair final state
of the Born and virtual processes. The spin-independent phase-space
integrals were first computated by Grunberg, Ng, and Tye for their study
of angular distributions of heavy-quark jets~\cite{grun}. In the spin-dependent
case, the obstacles to overcome are twofold: not only is the integral
more difficult to tackle, but also all symmetry properties of the
spin-independent integrals are lost. With a suitable choice of phase-space
parametrization in combination with sophisticated integration techniques, we
may derive the spin-dependent integrals in a closed analytical form~\cite{KPT}.
However, it is remarkable that all three-particle final-state phase-space
integrals relevant to the computation of the production cross sections take
a particularly simple form for certain limiting cases. In Tab.~1, we display
their limiting behavior for vanishing quark masses and as the
center-of-momentum
energy approaches the threshold value of $2m_q$.

A straightforward summation of the corresponding virtual and real
contributions yields the full analytical expressions for the factors
which the right hand sides of Eqs.~(\ref{sigvv})--(\ref{sigva}) must
be multiplied by to correct those cross sections to $O(\as)$
\bea
c^{VV} &=& 1 +{\as\over 2\pi}\,\CF\,\Bigg[\,
\left({1+v^2\over v}\,\ln{1-v\over 1+v}+2\right)
\ln\left(\Frac{1}{4}\xi\right)+F(v)
+v{1-v^2\over 3-v^2}\,\ln{1-v\over 1+v}\,\nn[.3cm]
&&
-{4\over v}\I{2}-{\xi\over v}\tilde{\I{3}}
+{4\over v(3-v^2)}\I{4}+{1+v^2\over v}\tilde{\I{5}}\,\bigg]
\label{cvv},\\[.6cm]
c^{AA} &=& 1 +{\as\over 2\pi}\,\CF\,\Bigg[\,
\left({1+v^2\over v}\,\ln{1-v\over 1+v}+2\right)
\ln\left(\Frac{1}{4}\xi\right)+F(v)
-2{1-v^2\over v}\,\ln{1-v\over 1+v}\nn[.3cm]
&&
+{\xi\over v^3}\I{1}-{4\over v}\I{2}-{\xi\over v}\tilde{\I{3}}
+{2+\xi\over v^3}\I{4}-{1+v^2\over v}\tilde{\I{5}}\,\Bigg], \\[.6cm]
c^{V\!A} &=& 1 +{\as\over 2\pi}\,\CF\,\Bigg[\,
\left({1+v^2\over v}\,\ln{1-v\over 1+v}+2\right)
\ln\left(\Frac{1}{4}\xi\right)+F(v)
-{1-v^2\over v}\,\ln{1-v\over 1+v}\nn[.3cm]
&&
+{1\over 2v^2}\bigg\{
(4-\xi)\S{1}-(4-5\xi)\S{2}-2(4-3\xi)\S{4}-\xi(1-\xi)(\tilde{\S{3}}
-\tilde{\S{5}}) \nn[.3cm]
&&
+\xi(\S{6}-\S{7})-2\S{8}+(2-\xi)\S{9}+(6-\xi)\S{10}
-2\S{11}+2(1+v^2)v^2\tilde{\S{12}}\,\bigg\}\Bigg],\ \ \label{cva}
\eea
where $\I{i}$ and $\S{i}$ are, respectively, the spin-independent and
spin-dependent integrals defined in Ref.~3. 
As these measures are
infrared safe, we have employed the notation $\tilde{\I{i}}$ and
$\tilde{\S{i}}$ for the finite part of the integrals $\I{i}$ and $\S{i}$,
{\it i.e.\/}, in Eqs.~(\ref{cvv})--(\ref{cva}) the soft divergences of the
virtual and real contributions have exactly canceled. The term $F(v)$ stems
from the virtual process and is given by
\be
F(v)\!\!=\!\!\left(\!3v\!-{1+v^2\over 2v}\ln{4v^2\over 1-v^2}\!\right)\!
\ln{1+v\over 1-v}+{1+v^2\over v}\!\real\!\!\left[
\Li\!\left({v+1\over 2v}\right)\!-\!\Li\!\left({v-1\over 2v}\right)\right]
+{\pi^2\over 2}{1+v^2\over v}-4
\ee
and as usual $\CF=4/3$ is the Casimir operator of the color group $SU(3)$.
It is easy to recognize that $c^{A\,V}=c^{V\!A}$ follows from the symmetry
properties of the fermionic trace structure.

In the high $(v\to 1)$ and low $(v\to 0)$ energy limits, the $O(\as)$
factors of Eqs.~(\ref{cvv})--(\ref{cva}) greatly simplify. With
the explicit results provided in Tab.~1, we can straightforwardly find
the following generic form
\be\label{cij}
c^{ij} = 1 + \CF\as \left[\,
{\pi\over 2v}-f^{ij}(v)\left({\pi\over 2}-{\rho^{ij}\over 4\pi}\right)\,\right]
\quad\mbox{with}\quad f^{ij}(1)\equiv 1,
\ee
where the functions $f^{ij}(v)$ and the constants $\rho^{ij}$ depend on the
specific parity-parity combination $i,j=V,A$. The constants $\rho^{ij}$ are
derived as
\be
\rho^{VV}=\rho^{AA}=3\qquad\mbox{and}\qquad\rho^{V\!A}=1,
\ee
which directly implies that the parity-even and parity-odd one-loop
corrections to the total correlation cross sections do not equal in the
fermionic zero-mass limit. In fact, the one-loop factor $c^{V\!A}$
receives a finite contribution from a residual coupling of left- and
right-handed helicity states of the polarized quark even in the limit
$m_q\to 0$. The physical origin of this $O(\as)$ effect is that transitions
between both helicity states are still allowed in the massless limit through
the emission of a real gluon.

The finite difference $\rho^{VV}-\rho^{V\!A}$ between the results of the
limiting case $m_q\to 0$ and the results where the fermion mass is at the
outset zero, is nothing more than a manifestation of the distinct nature of the
underlying chiral symmetries in strictly massless and massive theories.
Explicitly, we have
\be\label{anom}
\rho^{VV}-\rho^{V\!A} = \lim_{\xi\to 0}\;\xi\,\S{7} = 4\pi^2 R,
\ee
where $R=1/2\pi^2$ is the absorptive part of the axial anomaly in the limit
$m_q\to 0$ (the``anomaly pole''~\cite{ABJ,huang}\/). In the
fermionic zero-mass limit, the violation of chiral invariance
in triangle graphs with one axial-vector source is directly related to
the breaking of chiral invariance for transitions between fermions of
different helicity states~\cite{LN,DZ,SFS}.

We can connect the two boundary-value functions for vanishing quark mass
and for the center-of-momentum energy squared approaching $s=4m_q^2$
by simple interpolation formulae which are polynomials of degree $n$
\be
f(v) = {\displaystyle\sum\limits_{k=0}^n a_k v^k\Bigg/
\displaystyle\sum\limits_{k=0}^n a_k},
\ee
where $a_0,\ldots,a_n$ are suitable fitting constants. Our low-order
approximations for the $VV$ and $AA$ one-loop {\it QCD} corrections are
in agreement with the results which have been presented in the literature
before~\cite{schwinger,zerwas}
\bea\label{lo}
c^{VV} &=& 1 + \CF\as \left[\,{\pi\over 2v}-
{3+v\over4}\left({\pi\over2}-{3\over4\pi}
\right)\,\right], \\[.5cm]
c^{AA} &=& 1 + \CF\as \left[\,{\pi\over2v}-
{19-44v+35v^2\over10}\left({\pi\over2}-
{3\over4\pi}\right)\,\right].
\eea
The corresponding new formula that interpolates the parity-odd contribution
Eq.~(\ref{cva}) between the exact solution for $v=1$ and in the asymptotic
energy range near threshold is
\be\label{IP}
c^{V\!A} = 1 + \CF\as \left[\,{\pi\over 2v}-
{64-70v+103v^2\over 97}\left({\pi\over 2}-{1\over 4\pi}\right)\,\right].
\ee
In Fig.~1, we have plotted $c^{V\!A}(v)$ for a constant coupling $\as=0.1$
with the correct limit $c^{V\!A}(1)=1+\CF\as/4\pi$. The solid line gives
the analytical result of Eq.~(\ref{cva}), whereas the dashed line refers
to the interpolation formula Eq.~(\ref{IP}). Up to $v=0.7$, Eq.~(\ref{IP})
provides an excellent approximation for the $O(\as)$ correction that
multiplies with $\sigma^{V\!A}_S$.

For higher precision, we give in Tab.~2 polynomial approximations of all
possible parity-parity combinations $(ij)$ up to 5th order. The coefficients
$a^{ij}_k$ and $\rho^{ij}$ completely determine the structure of $c^{ij}$ in
Eq.~(\ref{cij}).

To demonstrate the accuracy of the entirely new Schwinger-type interpolation
formulae for the one-loop correction $c^{V\!A}$, we have plotted only the
corresponding non-trivial terms $f^{V\!A}(v)$ as a function of $v$. Fig.~2
displays the numerical results for the exact analytical form Eq.~(\ref{cva})
(solid line), for the low-order approximation Eq.~(\ref{IP}) (long-dashed
line),
and for the various polynomial interpolations presented in Tab.~2. We have
drawn the polynomials of 3rd order (short-dashed line), 4th order (dot-dashed
line) and 5th order (dotted line).

In Tab.~3, we present the $O(\as)$ numerical estimates for the total
unpolarized
rate $\sigma_{tot}=\sigma(e^+ e^-\to\gamma,Z\to c\bar{c}(g))$ and for the
longitudinal spin polarization $\left\langle\PL\right\rangle$ defined by
\be
\left\langle\PL\right\rangle =
{\sigma(\lambda_+)-\sigma(\lambda_-)\over\sigma_{tot}},
\ee
where the following shorthand notation has been used
\be
\sigma(\lambda_\pm)=\sigma(e^+ e^-\to\gamma,Z\to q(\lambda_\pm)\bar{q}(g))
=\pm v^2\,c^{V\!A}\,\sigma^{V\!A}_S.
\ee
In the calculation we have incorporated the running of the quark mass
and the $q^2$-evolution of the strong coupling according to
Refs.~14 
with $\as(\MZ^2)=0.12$ and
$\Lambda_{\overline{\scriptscriptstyle MS}}=0.238\,\mbox{\rm GeV}$,
where $\overline{MS}$ denotes the modified minimal subtraction scheme.
Column~I corresponds to the exact analytical expressions using
Eqs.~(\ref{cvv})--(\ref{cva}), whereas columns~II and III employ the
low-order approximation Eqs.~(\ref{lo})--(\ref{IP}) and the 3rd order
interpolation formulae $c^{ij}$ with the coefficients $a_k^{ij}$ of Tab.~2.
We find that already the low-order approximation provides a very accurate
interpolation to the exact analytical results and even for
$\sqrt{s}<100\,\mbox{\rm GeV}$ the higher-order approximations give only
little improvement. In Tab.~4 and Tab.~5, the exact results (I) are compared
with the low-order (II) and 5th order (III) approximations for bottom-
and top-quark production, respectively.

In the Standard Model the longitudinal spin polarization on the $Z$-peak
is fairly large yielding for the strictly massless Dirac theory a Born value of
$\left\langle\PL\right\rangle_{m_{c,t}=0}=-68.5\%$ (up-type quarks) and
$\left\langle\PL\right\rangle_{m_b=0}=-93.9\%$ (down-type quarks).
As pointed out before, at the one-loop order of quantum corrections a
correct theory for vanishing quark masses should always be regarded as
the limit of a massive theory, thus including the anomalous contribution
of Eq.~(\ref{anom}). With $R\neq0$ one obtains in this limit
\be
\left\langle\PL\right\rangle_{m_q\to 0}=
{\strut 1+\CF\as\,(\rho^{VV}-4\pi^2 R)/4\pi\over\strut
 1+\CF\as\;\rho^{VV}/4\pi}\,
\left\langle\PL\right\rangle_{m_q=0},
\ee
which gives a sizable effect of the order of $3\%$. We anticipate that
future experimental analyses of $\left\langle\PL\right\rangle$ via the
decay products of the final quark pair will be able to detect this effect.
The angular distributions of the charged leptons from semileptonic
decays of the charmed Lambda~\cite{andrzei} would serve as spin analyzers for
the heavy quark, since in the heavy-quark limit the polarization information of
the quark $q$ is completely transferred to the corresponding Lambda baryon
$\Lambda_q$~\cite{CKPS}.

To summarize, we have presented the analytical expressions for the total
unpolarized and polarized production cross sections for the process
$e^+ e^-\to\gamma,Z\to q\bar{q}$ up to one-loop order of strong interactions.
The numerical results for the total rate and longitudinal spin polarization
are compared with estimates Schwinger-type interpolation
formulae give. The simple second-order approximation for the parity-odd
correlation cross section already yields very accurate results. Use of
this compact formula may be expected to provide accurate numerical estimates
in an uncomplicated way.

In this context, we have also pointed out that finite contributions from
the axial anomaly are relevant in the limit of vanishing quark mass.
The breaking of chiral invariance for transitions between fermions of
different helicity states fundamentally alters the high-energy behavior
of parity-odd observables, such as the longitudinal polarization asymmetry.
For heavy-quark production, these chirality-violating mass effects reduce the
longitudinal polarization by approximately $3\%$ and thus should be observable
at future TeV $e^+ e^-$ colliders by analyzing the charged lepton spectrum
of the semileptonic decay mode.

\newpage\noindent
{\bf Acknowledgements.} This research was supported by the German Academic
Exchange Program and in part by the CICYT, Spain, under Grant AEN 93/0234.
I am particularly grateful to A.~Pilaftsis for many stimulating discussions.
It is also a pleasure to thank J.~Bernab\'eu, J.G.~K\"orner, B.~Lampe,
M.~Lavelle, and J.~Pe\~narrocha for helpful conversations.

\newpage

\centerline{\bf\Large Table Captions }
\vspace{1cm}
\newcounter{tab}
\begin{list}{\bf\rm Tab.\ \arabic{tab}: }{\usecounter{tab}
\labelwidth1.6cm \leftmargin2.5cm \labelsep0.4cm \itemsep0ex plus0.2ex }

\item The behavior of the spin-independent $(I_i)$ and spin-dependent
      $(S_i)$ phase-space integrals for the decay $Z,\gamma\to q\bar{q}g$
      in the limit of vanishing quark mass $(v\to 1)$ and in the asymptotic
      energy range near threshold $(v\to 0)$. The zeta function of order 2
      is $\zeta(2)=\pi^2/6$.

\item One-loop {\it QCD} corrections for the
      individual parity-parity combinations in the
      process $e^+ e^-\to \gamma, Z\to q\bar{q}$.
      The functions $f^{ij}(v)$ give the non-trivial ingredient
      in the interpolation formulae for the total correlation cross
      sections.

\item $O(\as)$ production cross section and longitudinal polarization asymmetry
      for the process $e^+ e^- \to\gamma,Z\to c\bar{c}$ as a function of
      the c.m.s energy. Column~I gives the exact numerical results,
      whereas columns~{II} and {III} use the low-order approximation
      Eqs.~(\ref{lo})--(\ref{IP}) and the 3rd order interpolation formulae
      of Tab.~2, respectively.

\item $O(\as)$ production cross section and longitudinal polarization
      for the process $e^+ e^- \to\gamma,Z\to b\bar{b}$; column~I:
      exact results; column~{II}: low-order approximation; column~{III}:
      5th order interpolation.

\item $O(\as)$ production cross section and longitudinal polarization
      for the process $e^+ e^- \to\gamma,Z\to t\bar{t}$ with
      $m_t=174\mbox{\rm GeV}$;
      column~I: exact results; column~{II}: low-order approximation;
      column~{III}: 5th order interpolation.

\end{list}
\newpage

\centerline{\bf\Large Figure Captions }
\vspace{1cm}
\newcounter{fig}
\begin{list}{\bf\rm Fig.\ \arabic{fig}: }{\usecounter{fig}
\labelwidth1.6cm \leftmargin2.5cm \labelsep0.4cm \itemsep0ex plus0.2ex }

\item The $O(\as)$ correction factor
$c^{V\!A}=\sigma^{V\!A}_S/\sigma^{V\!A}_S(Born)$ given as a function of
$v=\sqrt{1-4m_q^2/s}$. The solid line represents the correct numerical
values, whereas the dashed line corresponds to the low-order Schwinger-type
interpolation formula Eq.~(\ref{IP}).

\item The $v$-dependence of the non-trivial term $f^{V\!A}(v)$ in the
one-loop {\it QCD} vector axial-vector correction factor $c^{V\!A}$ as
defined in Eq.~(\ref{cij}). Shown are the exact result Eq.~(\ref{cva})
(solid line), the low-order interpolation Eq.~(\ref{IP}) (long-dashed line),
and the polynomial approximations of 3rd (short-dashed line), 4th
(dot-dashed line) and 5th (dotted line) order given in Tab.~2.

\end{list}

\newpage
\begin{thebibliography}{99}

\bibitem{LEP} The LEP Collaborations and LEP Electroweak Working Group,
CERN-PPE/94-100, and references therein.

\bibitem{SLAC} The SLD and SLC Collaborations,
SLAC-PUB-6313, August 1993, and references therein.

\bibitem{KPT} J.G.\ K\"orner, A.\ Pilaftsis, and M.M.\ Tung,
Z.\ Phys.\ {\bf C63} (1994) 575; M.M.~Tung, Ph.D.
thesis, University of Mainz, 1993.

\bibitem{jers} J.\ Jersak, E.\ Laermann, and P.M.\ Zerwas,
Phys.\ Rev.\ {\bf D25} (1982) 363; A.\ Djouadi, J.H.\ K\"uhn and
P.M.\ Zerwas, Z.\ Phys.\ {\bf C46} (1990) 411.

\bibitem{alex} J.\ Alexander {\it et.\ al.\/}, Phys.\ Rev.\ {\bf D37}
(1988) 56.

\bibitem{grun} G.~Grunberg, Y.J.~Ng, and S.-H.H.\ Tye, Phys.\ Rev.\
{\bf D21} (1980) 62.

\bibitem{ABJ} S.~Adler, Phys.\ Rev.\ {\bf 177} (1969) 2426;
J.S.~Bell, and R.~Jackiw, Nuovo Cimento {\bf A60} (1969) 47.

\bibitem{huang} K.\ Huang, {\it Quarks, Leptons and Gauge Fields}
(World Scientific, Singapore, 1982), pp.\ 228--232, and references therein.

\bibitem{LN} T.D.~Lee and M.~Nauenberg, Phys.\ Rev.\ {\bf B133} (1964) 1594.

\bibitem{DZ} A.D.\ Dolgov and V.I.\ Zakharov, Nucl.\ Phys.\ {\bf B27}
(1971) 525.

\bibitem{SFS} A.V.~Smilga, Commun.\ Nucl.\ Part.\ Phys.\ {\bf 20} (1991) 69;
B.~Falk and L.M.~Sehgal, Phys.\ Lett.\ {\bf B325} (1994) 509.

\bibitem{schwinger} J.\ Schwinger, {\it Particles, Sources and Fields}
(Addison-Wesley, Redwood City, 1988), {\bf Vol.\ III}, pp.\ 99 and
pp.\ 109.

\bibitem{zerwas} P.M.\ Zerwas, {\it $e^+ e^-$ Linear Colliders: Physics
Prospects\/}, published in EFCA LC 1992: 11--92 (QCD183:I78:1992).

\bibitem{GL} J.\ Gasser and H.\ Leutwyler, Phys.\ Rep.\ {\bf 87} (1982) 77;
K.\ Kang, J.\ Flanz, and E.\ Paschos, Z.\ Phys.\ {\bf C55} (1992) 75.

\bibitem{andrzei} A.\ Czarnecki and M.\ Je\.zabek, Nucl.\ Phys.\
{\bf B427} (1994) 3, and references therein.

\bibitem{CKPS} F.E.\ Close, J.G.\ K\"orner, R.J.N.\ Phillips, and
D.J.\ Summers, J.\ Phys.\ {\bf G18} (1992) 1716.

\end{thebibliography}
\end{document}